\documentclass[preprint,showpacs,preprintnumbers,amsmath,amssymb]{revtex4}

\usepackage{graphicx}
\usepackage{epsfig}		
\usepackage{dcolumn}
\usepackage{bm}

\def\0{\mbox{\tiny $0$}}
\def\1{\mbox{\tiny $1$}}
\def\2{\mbox{\tiny $2$}}
\def\3{\mbox{\tiny $3$}}
\def\4{\mbox{\tiny $4$}}
\def\5{\mbox{\tiny $5$}}
\def\6{\mbox{\tiny $6$}} 
\def\7{\mbox{\tiny $7$}}
\def\8{\mbox{\tiny $8$}}
\def\9{\mbox{\tiny $9$}}
\def\R{\mbox{\tiny $R$}}
\def\T{\mbox{\tiny $T$}}

\def\I{\mbox{\tiny $I$}}

\def\k{\mbox{\tiny $k$}}
\def\w{\mbox{\tiny $w$}}
\def\L{\mbox{\tiny $L$}}

\def\I{\mbox{\tiny $I$}}
\def\D{\mbox{\tiny $D$}}
\def\mi{\mbox{\tiny $-$}}
\def\pl{\mbox{\tiny $+$}}

\begin{document}

\title{THE EXACT CORRESPONDENCE BETWEEN PHASE TIMES AND DWELL TIMES IN A SYMMETRICAL QUANTUM TUNNELING CONFIGURATION}

\author{A. E. Bernardini}
\email{alexeb@ifi.unicamp.br}
\affiliation{Instituto de F\'{\i}sica Gleb Wataghin, UNICAMP,
PO Box 6165, 13083-970, Campinas, SP, Brasil.}

\date{\today}

\begin{abstract}
The general and explicit relation between phase times and dwell times for quantum tunneling or scattering is investigated.
Considering two identical propagating wave packets symmetrically impinging a one-dimensional barrier, here we demonstrate that these two distinct transit time definitions give connected results where, for such a colliding configuration, the phase time (group delay) accurately describes the exact position of the scattered particles.
The analytical difficulties that arise when the stationary phase method is employed for obtaining the phase (traversal) times are all overcome since the multiple wave packet decomposition allows us to recover the exact position of the reflected and transmitted waves.
In addition to the exact relation between the phase time and the dwell time, leads to right interpretation for both of them.
\end{abstract}

\date{\today}
\maketitle

Finding a definitive answer for the time spent by particle to penetrate a classically forbidden region delimited by a potential barrier has occupied the physicists for decades \cite{PBE,Con70,But83,Hau87,Fer90,Bro94,Sok94,Olk95,Jak98,Olk02,WinWin,Olk04}.
In particular, people have tried to introduce quantities that have dimension of time and can somehow be associated with the passage of the particle through the barrier or, strictly speaking, with the definition of the tunneling time.
These proposals have led to the introduction of several {\em transit time} definitions that can be summarized by three groups.
(1) The first group comprises a time-dependent description in terms of wave packets where some features of an incident packet and the comparable features of the transmitted packet are 
utilized to describe a quantifiable {\em delay} as a tunneling time \cite{Hau89}.
(2) In the second group the tunneling times are computed based on averages over a set of kinematical paths, whose distribution is supposed to describe the particle motion inside a barrier.
In this case, Feynman paths are used like real paths to calculate an average tunneling time with the weighting function $\exp{[i\, S\, x(t)/\hbar]}$, where $S$ is the action associated with the path $x(t)$ (where $x(t)$ represents the Feynman paths initiated from a point on the left of the barrier and ending at another point on the right of it \cite{Sok87}).
The Wigner distribution paths \cite{Bro94}, and the Bohm approach \cite{Ima97,Abo00} are included in this group.
(3) In the third group we notice the introduction of a new degree of freedom, constituting a physical clock for the measurements of tunneling times.
This group comprises the methods with a Larmor clock \cite{But83} or an oscillating barrier \cite{But82}.
Separately, standing on itself is the {\em dwell} time defined by the interval during which the incident flux has to exist and act, to provide the expected accumulated particle storage, inside the barrier \cite{Lan94}.
In spite of no general agreement \cite{Olk04,Olk92} among the above definitions,
the so called phase time \cite{Wig55} (group delay) and the dwell time have an apparently well established relation between them \cite{Hau89,Win03}.
However, these time definitions remain controversial since in the opaque barrier limit
they predict effective tunneling velocities that exceed the vacuum speed of light and may even become unlimited (Hartman effect)\cite{Har62}.
For instance, some of the barrier traversal time definitions lead, under tunneling conditions, to very short times, which can even become negative.
It can precipitately induces an interpretation of violation of simple concepts of causality.
Otherwise, negative speeds do not seem to create problems with causality, since they were predicted both within special relativity and within quantum mechanics \cite{Olk95}.
A possible explanation for such time advancements can come, in any case, from consideration of the very rapid spreading of the initial and transmitted wave packets for large momentum distribution widths.
Due to the similarities between tunneling (quantum) packets and evanescent (classical) waves, exactly the same phenomena are to be expected in the case of classical barriers\footnote{In particular, we could mention the analogy between the stationary Helmholtz equation for an electromagnetic wave packet - in a waveguide, for instance - in the presence of a {\em classical} barrier and the stationary Schroedinger equation, in the presence of a potential barrier \cite{Lan94,Jak98,Nim94}).}.
The existence of such negative times is predicted by relativity itself based on its ordinary postulates \cite{Olk04}, and they appear to have been experimentally detected in many works \cite{Gar70,Chu82}.
In some recent analysis, the proportionality between the phase time and the time averaged stored energy was used to explain these unusual effects \cite{WinWin, Win03}.
In particular, from the time-independent Schroedinger equation, a relation between the group delay and the dwell time was derived for quantum tunneling .

In this manuscript we are concentrated on overcoming the difficulties that appear when the stationary phase method (SPM) is utilized for deriving tunneling times in order to accurately investigate the relation between the phase time delay and the dwell time for quantum tunneling or scattering. 
Taking into account the restrictive conditions for the use of the method, we report about a theoretical configuration involving a symmetrical collision between two identical wave packets and a one-dimensional rectangular potential barrier \cite{Ber06}.
Using the procedure we call multiple peak decomposition \cite{Ber04}, we demonstrate that summing the amplitudes of the reflected and transmitted waves allows the reconstruction of the scattered wave packets so that the analytical conditions for the stationary phase principle applicability can be totally recovered.
For such a colliding configuration, the phase time gives the exact position of the {\em center of mass} of each symmetrically scattered wave packet.
Consequently, we can have a realistic idea of the {\em magnitude} of the dwell time for the case in which the energy of the incident particle is smaller than the barrier potential energy (tunneling configuration).
In spite of the theoretical focus, the results here obtained apply only to such configurations which should deserve further attention by experimenters, while the existing experiments report the inaccurate results on non-symmetrical configurations.

In general lines, the SPM allows one to describe the movement of the center of a wave packet constructed in terms of a momentum distribution $g(k - k_{\0})$ which has a pronounced peak around $k_{\0}$. 
Assuming that the characteristic phase of the propagation varies smoothly around the maximum of $g(k - k_{\0})$, the stationary phase condition enables us to calculate the position of the peak of the wave packet (highest probability region to find the propagating particle).
With regard to the {\em standard} one-way direction wave packet tunneling,
in which we consider a rectangular potential barrier $V(x)$, $V(x) = V_{\0}$ if $x \in \mbox{$\left[- L/2, \, L/2\right]$}$
and $V(x) = 0$ if $x \in\hspace{-0.3cm}\slash\hspace{0.1cm}\mbox{$\left[- L/2, \, L/2\right]$}$,
it is well-known \cite{Ber06} that the transmitted amplitude $T(k, L) = |T(k, L)|\exp{[i \Theta(k, L)]}$ is written in terms of
\small\begin{equation}
|T(k, L)| =
\left\{1+ \frac{w^4}{4 \, k^{\2} \, \rho^{\2}(k)}
\sinh^{\2}{\left[\rho(k)\, L \right]}\right\}^{-\frac{1}{2}},
\label{501}
\end{equation}\normalsize
and
\small\begin{equation}
\Theta(k, L) = \arctan{\left\{\frac{2\, k^{\2} - w^{\2}}
{2\,k \, \rho(k)}
\tanh{\left[2\,\rho(k) \, L \right]}\right\}}, 
\label{502}
\end{equation}\normalsize
for which we have made explicit the dependence on the barrier length $L$, and we have adopted $\rho(k) = \left(w^{\2} - k^{\2}\right)^{\frac{1}{2}}$ with $w = \left(2\, m \,V_{\0}\right)^{\frac{1}{2}}$ and $\hbar = 1$.
The above result is adopted for calculating the transit time $t_{T}$ of a transmitted wave packet when its peak emerges at $x = L/2$,
\small\begin{equation}
t_{T} =\frac{m}{k_{\0}}\left.\frac{d\Theta(k, \alpha)}{dk}\right|_{_{k = k_{\0}}} =
\frac{2\,m \, L}{k_{\0} \,\alpha }
\left\{\frac{w^4\,\sinh{(\alpha)}\cosh{(\alpha)}
-\left(2\, k_{\0}^{\2} - w^{\2} \right)k_{\0}^{\2} \,\alpha }
{4\, k_{\0}^{\2} \,\left(w^{\2} - k_{\0}^{\2} \right)  +
w^4\,\sinh^{\2}{(\alpha)}}\right\}.
\label{4}
\end{equation}\normalsize
Here we have defined the parameter $\alpha = w L \sqrt{(1 - k_{\0}^{\2}/w^{\2})}$
and $k_{\0}$ as the maximum of a generic symmetrical momentum distribution $g(k - k_{\0})$ that composes the {\em incident} wave packet.
By following our previous analysis \cite{Ber06}, it is well-established that, due to the {\em filter effect}, the amplitude of the transmitted wave is essentially composed by the plane wave components of the front tail of the {\em incoming} wave packet which reaches the first barrier interface before the peak arrival \cite{Lan89}.
We have shown that the {\em cut off} of the momentum distribution at $k \approx (1 - \delta) w$ increases the amplitude of the tail of the incident wave so that it contributes so relevantly as the peak of the incident wave to the final composition of the transmitted wave.
Independently, due to the novel asymmetric character of the transmitted amplitude $g(k-k_{\0}) |T(k, L)|$, an ambiguity in the definition of the {\em arrival}/{\em transmitted} time is created \cite{Ber06}.
In the framework of the multiple peak decomposition \cite{Ber04}, we have suggested a suitable way for comprehending the conservation of probabilities
where the asymmetric aspects previously discussed \cite{Ber06} could be totally eliminated.
In order to recover the scattered momentum distribution symmetry conditions for accurately applying the SPM, we assume a symmetrical colliding configuration of two wave packets traveling in opposite directions. 
By considering the same rectangular barrier $V(x)$, we solve the Schroedinger equation for a plane wave component of momentum $k$ for two identical wave packets symmetrically separated from the origin $x = 0$.
By assuming that $\phi^{\L(\R)}(k,x)$ are Schroedinger equation stationary wave solutions,
when the wave packet peaks simultaneously reach the barrier (at the mathematically convenient time $t = - (m L) /(2 k_{\0})$) we can write
\small\begin{equation}
\phi^{\L(\R)}(k,x)=
\left\{\begin{array}{l l l l}
\phi^{\L(\R)}_{\1}(k,x) &=& 
\exp{\left[ \pm i \,k \,x\right]} + R^{\L(\R)}(k,L)\exp{\left[ \mp i \,k \,x\right]}&~~~~x < - L/2\, (x > L/2),\nonumber\\
\phi^{\L(\R)}_{\2}(k,x) &=& \gamma^{\L(\R)}(k)\exp{\left[ \mp\rho  \,x\right]} + \beta^{\L(\R)}(k)\exp{\left[ \pm\rho  \,x\right]}&~~~~- L/2 < x < L/2,\nonumber\\
\phi^{\L(\R)}_{\3}(k,x) &=& T^{\L(\R)}(k,L)\exp{ \left[\pm i \,k \,x\right]}&~~~~x > L/2 \, (x < - L/2) .
\end{array}\right.
\label{510}
\end{equation}\normalsize
where the upper(lower) sign is related to the index $L$($R$) corresponding to the incidence on the left(right)-hand side of the barrier.
By assuming the conditions for the continuity of $\phi^{\L,\R}$ and their derivatives at $x = - L/2$ and $x = L/2$, after some mathematical manipulations, we can easily obtain
\small\begin{equation}
R^{\L,\R}(k,L) = \exp{\left[ - i \,k \,L \right]} \left\{\frac{\exp{\left[ i \, \theta(k)\right]} \left[1 - \exp{\left[ 2\,\rho(k) \,L\right]}\right]}{1 - \exp{\left[ 2\,\rho(k) \,L\right]}\exp{\left[ i\, 2\,\theta(k)\right]}}\right\}
\label{511}
\end{equation}\normalsize
and
\small\begin{equation}
T^{\L,\R}(k,L) = \exp{\left[ - i \,k \,L \right]} \left\{\frac{\exp{\left[\rho(k) \,L\right]}\left[1- \exp{\left[ 2\, i\, \theta(k)\right]}\right]}{1 - \exp{\left[ 2\,\rho(k) \,L\right]}\exp{\left[ i\, 2\,\theta(k)\right]}}\right\},
\label{512}
\end{equation}\normalsize
where
\small\begin{equation}
\theta(k) = \frac{ 2\, k \, \rho(k)}{2k^{\2} - w^{\2}}
\label{512B}
\end{equation}\normalsize
and $R^{\L(\R)}(k,L)$ and $T^{\R(\L)}(k,L)$ are intersecting each other.

Since the above colliding configuration is spatially symmetric, the symmetry operation corresponding to the $1 \leftrightarrow 2$ (or $L \leftrightarrow R$) particle exchange can be parameterized by the position coordinate transformation $x \rightarrow -x$.
At the same time, taking advantage from the notation that we have adopted, it is easy to observe that
\small\begin{equation}
\phi^{\L(\R)}(k,x) = \phi^{\L(\R)}_{\1 \pl \2 \pl \3}(k,x) = \phi^{\R(\L)}_{\1 \pl \2 \pl \3}(k,-x) = \phi^{\R(\L)}(k,-x) 
\label{512C}
\end{equation}\normalsize
where the $L \leftrightarrow R$ interchange is explicit. 
Consequently, in case of analyzing the collision of two identical bosons, we have to consider a symmetrized superposition of the $L$ and $R$ wave functions,
\small\begin{equation}
\phi_{\pl}(k,x) = \phi^{\L}(k,x) + \phi^{\R}(k,x) = \phi^{\R}(k,-x) + \phi^{\L}(k,-x) = \phi_{\pl}(k,-x).
\label{512D}
\end{equation}\normalsize
Analogously, in case of analyzing the collision of two identical fermions (just taking into account the spatial part of the wave function),
we have to consider an antisymmetrized superposition of the $L$ and $R$ wave functions,
\small\begin{equation}
\phi_{\mi}(k,x) = \phi^{\L}(k,x) - \phi^{\R}(k,x) = \phi^{\R}(k,-x) - \phi^{\L}(k,-x) = -\phi_{\mi}(k,-x).
\label{512E}
\end{equation}\normalsize
Thus the amplitude of the re-composed transmitted plus reflected waves would be given by
$R^{\L,\R}(k,L) + T^{\R,\L}(k,L)$ for the symmetrized wave function $\phi_{\pl}$ and by
$R^{\L,\R}(k,L) - T^{\R,\L}(k,L)$ for the antisymmetrized wave function $\phi_{\mi}$.
Reporting to the previously introduced procedure \cite{Ber04} that we call multiple peak decomposition, for such a pictorial symmetrical tunneling configuration, we can superpose the amplitudes of the intersecting probability distributions before taking their squared modulus in order to obtain
\small\begin{eqnarray}
R^{\L,\R}(k,L) \pm T^{\R,\L}(k,L) &=& \exp{\left[ - i \,k \,L \right]} \left\{\frac{\exp{\left[\rho(k) \,L\right]}\pm \exp{\left[ i\, \theta(k)\right]}}{1 \pm \exp{\left[ \rho(k) \,L\right]}\exp{\left[ i\, \theta(k)\right]}}\right\}\nonumber\\
&=& \exp{\left\{ - i [k \,L - \varphi_{\pm}(k,L)]\right\}}
\label{513}
\end{eqnarray}\normalsize
with
\small\begin{equation}
\varphi_{\pm}(k,L) = - \arctan{\left\{\frac{2\,k\,\rho(k) \, \sinh{[\rho(k)\,L]}}{\left(k^{\2}-\rho^{\2}(k)\right)\cosh{[\rho(k)\,L]} \pm w^{\2}}\right\}}.
\label{514}
\end{equation}\normalsize\normalsize
where the {\em plus} sign is related to the results obtained for the symmetrized superposition and the {\em minus} sign is related to the antisymmetrized superposition.
Independently of the symmetrization characteristic of the wave function, we observe from Eq.~(\ref{513}) that, differently from the previous standard tunneling analysis, by adding the intersecting amplitudes at each side of the barrier, we keep the original momentum distribution undistorted since $|R^{\L,\R}(k,L) \pm T^{\R,\L}(k,L)|$ is equal to one.
At this point we recover the most fundamental condition for the applicability of the SPM .
It allows us to accurately find the position of the peak of the reconstructed wave packet composed by reflected and transmitted superposing components.
The phase time interpretation can be, in this case, correctly quantified in terms of the analysis of the novel phase $\varphi_{\pm}(k, L)$.

From this point we have opted for performing the analysis of the symmetrized superposition from which, for simplicity, we have suppressed the {\em plus} sign from the notation.
We firstly notice that the novel scattering amplitudes $g(k - k_{\0})|R^{\L,\R}+ T^{\R,\L}| \approx g(k - k_{\0})$ still remain symmetric and, from the SPM, the scattering phase time, results in
\small\begin{equation}
t^{(\alpha)}_{T, \varphi} =\frac{m }{k}\frac{d\varphi(k, \alpha)}{dk} =
\frac{2\,m\, L}{k\,\alpha}
\frac{w^{\2}\sinh{(\alpha)} + \alpha\,k^{\2}}{2\,k^{\2} - w^{\2} + w^{\2}\cosh{(\alpha)}}
\label{515}
\end{equation}\normalsize
where $k \rightarrow k_{\0}$, with $\alpha$ previously defined \footnote{A similar procedure can be carried out for the antisymmetrized wave function configuration.}. 
Hereafter, we cannot differentiate the tunneling from the reflecting waves for such a scattering configuration.
We have simply demonstrated that the transmitted and reflected interfering amplitudes results in a unimodular function which just modifies the {\em envelop} function $g(k - k)$ by an additional phase.
The previously pointed out incongruities which cause the distortion of the momentum distribution $g(k - k)$ are completely eliminated in this case.
The point is that the {\em old} phase $\Theta(k, L)$ (Eq.~\ref{502}) appears when we treat separately the momentum amplitudes
$T(k, L)$ and $R(k, L)$, which destroys the symmetry of the initial momentum distribution $g(k - k_{\0})$ by the presence of the
multiplicative term $T(k, L)$ or $R(k, L)$, and the novel phase $\varphi(k, L)$ appears only when we sum the tunnneling/scattering amplitudes so that the symmetrical character of the initial momentum distribution is maintained (due to the result of Eq.~(\ref{513})).
It {\em requalifies} the SPM for accurately computing the time dependence of the position of the peak of a wave packet. 
At the same time, one could argue about the possibility of extending such a result to the tunneling process established in a standard way.
We should assume that in the region inside the potential barrier, the reflected and transmitted amplitudes should be summed before we compute the phase changes.
Obviously, it would result in the same phase time expression as represented by (\ref{515}).
In this case, the assumption of there (not) existing interference between the momentum amplitudes of the reflected and transmitted waves at the discontinuity points $x = -L/2$ and $x = L/2$ is purely arbitrary.
Consequently, it is important to reinforce the argument that such a possibility of interference leading to different phase times
is strictly related to the idea of using (or not) the multiple peak (de)composition in the region where the potential barrier is localized.  
\begin{figure}[th]
\vspace{-0.6 cm}
\centerline{\psfig{file=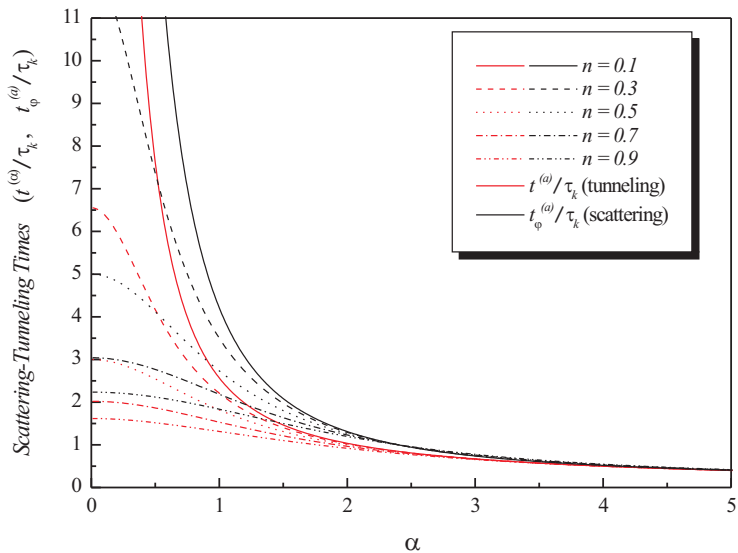,width=11cm}}
\vspace{-1 cm}
\caption{Normalized times for the {\em standard} one-way direction tunneling and the {\em symmetrical} scattering process.
These times can be understood as transit times in the units of the {\em classical} traversal time $\tau_{\k} = (m L) /k$.
Both present the same asymptotic behavior.
\label{fig3A}}
\vspace{-0.4 cm}
\end{figure}
To illustrate the difference between the {\em standard} tunneling phase time $t^{(\alpha)}_{T}$ and the {\em symmetrical} scattering phase time $t^{(\alpha)}_{T, \varphi}$ we introduce the parameter $n = k^{\2}/w^{\2}$ and we define the {\em classical} traversal time $\tau_{\k} = (m L) /k$.
The normalized phase times can then be written as
\small\begin{equation}
t^{(\alpha)}_{T}
= 
\frac{2  \tau_{\k}}{\alpha}\left\{
\frac{\cosh{(\alpha)}\sinh{(\alpha)} - \alpha\,n\left(2 n - 1\right)}{\left[4 n \left(1 - n\right)+\sinh^{\2}{(\alpha)}\right]}
\right\}
~~\mbox{and}~~
t^{(\alpha)}_{T, \varphi}
= 
\frac{2  \tau_{\k}}{\alpha}\left\{\frac{n\, \alpha + \sinh{(\alpha)}}{2n - 1 +\cosh{(\alpha)}}
\right\}\label{517}
\end{equation}\normalsize
which are plotted in Fig.(\ref{fig3A}) for some discrete values of $n$.

\begin{figure}[th]
\vspace{-0.6 cm}
\centerline{\psfig{file=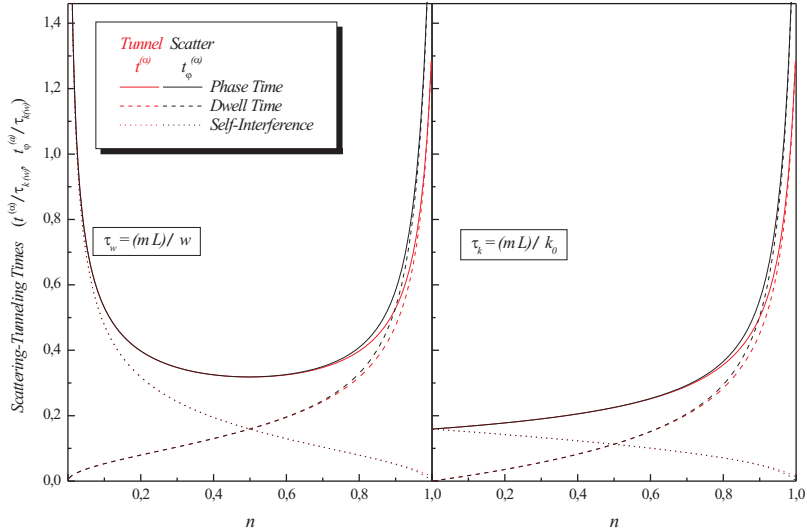,width=12cm}}
\vspace{-1 cm}
\caption{Exact phase time (solid line), self-interference delay (dotted line), and the dwell time (dashed line)
as a function of the normalized energy $n = k^{\2}/w^{\2} \propto E_{\0}/V_{\0}$ for the symmetrical (black line) and the standard one-way direction (red line) configuration
for wave packets colliding with a rectangular potential barrier. 
These times are normalized by the {\em barrier} time $\tau_{\w} = (m L) /w$ in the first plot and by the
{\em classical} traversal time $\tau_{\k} = (m L) /k$ in the second plot.
Here we have adopted $w L = 4\pi$ for $\alpha = w L\sqrt{1-n}$.
\label{fig2}}
\vspace{-0.4 cm}
\end{figure}

At this point, could one say metaphorically that the identical particles
represented by both impinging wave packets spend a time equal to $ t_{T, \varphi}$
inside the barrier before retracing its steps or tunneling?
The answer is in the definition of the dwell time for the same symmetrical colliding configuration that we have proposed above.
The dwell time is a measure of the time spent by a particle in the barrier region regardless of whether it is ultimately transmitted or reflected \cite{But83},
\small\begin{equation}
t_{D} 
=\frac{m}{k} \int_{\mi \L/\2}^{{\pl \L/\2}}\mbox{d}x{|\phi_{\2}(k,x)|^{\2}}
\label{530}
\end{equation}\normalsize
where  $j_{in}$ is the flux of incident particles and $\phi_{\2}(k,x)$ is the stationary state wave function depending on the colliding configuration that we are considering (symmetrical or standard).
To derive the relation between the dwell time and the phase time, we reproduce the variational theorem which yields the sensitivity
of the wave function to variations in energy.
After some elementary manipulations of the Schroedinger equation \cite{Smi60}, we can write
\small\begin{equation}
\phi^{\dagger}\phi = \frac{1}{2m}\frac{\partial}{\partial x}\left(\frac{\partial \phi}{\partial E}\frac{\partial \phi^{\dagger}}{\partial x} - \phi^{\dagger}\frac{\partial^{\2}\phi}{\partial E\partial x}\right).
\label{531}
\end{equation}\normalsize
Upon integration over the length of the barrier we find
\small\begin{equation}
2 m \int_{\mi \L/\2}^{{\pl \L/\2}}\mbox{d}x{|\phi_{\2}(k,x)|^{\2}} = \left.\left(\frac{\partial \phi}{\partial E}\frac{\partial \phi^{\dagger}}{\partial x} - \phi^{\dagger}\frac{\partial^{\2}\phi}{\partial E\partial x}\right)\right|_{\mi\L/\2}^{\pl\L/\2}.
\label{532}
\end{equation}\normalsize
In the barrier limits ($x = \pm L/2$), for the symmetrical configuration that we have proposed, we can use the superposition of
the scattered waves to substitute in the right-hand side of the above equation,
\small\begin{eqnarray}
\left.\phi(k,x)\right|_{\mi\L/\2(\pl\L/\2)} &=& \frac{\phi^{\L(\R)}_{\1}(k,x) + \phi^{\R(\L)}_{\3}(k,x)}{\sqrt{2}}\nonumber\\
&=&\exp{\left[ \pm i \,k \,x\right]} + \exp{\left[ \mp i \,k \,x + i \left(\varphi(k,L)- k L\right)\right]} 
\label{533}
\end{eqnarray}\normalsize
By evaluating the right-hand side of the Eq.~(\ref{533}), we obtain
\small\begin{equation}
\frac{\partial k}{\partial E} \frac{d\varphi}{dk} = \frac{m}{k} \int_{\mi \L/\2}^{{\pl \L/\2}}\mbox{d}x{|\phi_{\2}(k,x)|^{\2}}
- \frac{Im[\exp{(i \varphi)}]}{k} \frac{\partial k}{\partial E}.
\label{534}
\end{equation}\normalsize
The first term of the above equation is the phase time or the aforementioned group delay $t^{(\alpha)}_{T, \varphi}$.
The second term leads to the explicit computation of the dwell time.
By respecting the continuity conditions of the Schroedinger equation solutions,
in the barrier region we obtain a stationary wave symmetrical in $x$,
\small\begin{eqnarray}
\phi_{\2}(k,x) &=& \frac{\phi^{\L}_{\2}(k,x) + \phi^{\R}_{\2}(k,x)}{\sqrt{2}},~~~~~~(\gamma\equiv\gamma^{\L,\R}, \, \beta\equiv\beta^{\L,\R})\nonumber\\
&=& \sqrt{2}(\beta + \gamma)\cosh{[\rho(k)\,x]}, 
\label{534B}
\end{eqnarray}\normalsize
which, from Eq.~(\ref{530}), leads to
\small\begin{equation}
t^{(\alpha)}_{D, \varphi} =  
\frac{2\, \tau_{\k}\, n}{\alpha}
\frac{\sinh{(\alpha)} + \alpha}{2n - 1 + \cosh{(\alpha)}}.
\label{535}
\end{equation}\normalsize
The self-interference term which comes from the momentary overlap of incident and reflected waves in front of the barrier is given by\footnote{We have printed the phase index $\varphi$ for all the results related to the symmetrical colliding configuration.}
\small\begin{equation}
t^{(\alpha)}_{\I, \varphi} = - \frac{Im[\exp{(i \varphi)}]}{k} \frac{\partial k}{\partial E} = \frac{m \, \sin{(\varphi)}}{k^{\2}}=
\frac{2 \tau_{\k}}{\alpha}
\frac{(1-n)\sinh{(\alpha)}}{2n - 1 + \cosh{(\alpha)}}.
\label{536}
\end{equation}\normalsize
In conclusion, the dwell time is obtained from a simple subtraction of the quote self-interference delay
$t_{\I, \varphi}$ from the phase time that describes the exact position of the peak of the scattered wave packets,
$t_{\T,\varphi} = t_{\D, \varphi}+ t_{\I, \varphi}$,
as we can notice in Fig.\ref{fig2}.
Adopting the {\em classical} traversal time $\tau_{\k} = (m L) /k$ in place of the {\em barrier}
time $\tau_{\w} = (m L) /w = \tau_{\k} \sqrt{n}$ for normalizing the results illustrated in Fig.\ref{fig2}
may clarify some important aspects.
Let us assume that the amplitude of each separated transmission coefficient $T$ prevails over the amplitude of each
reflection coefficient $R$, i. e. $|T|^{\2} > |R|^{\2} \longrightarrow |T|^{\2} > 1/2$.
For satisfying such a requirement the Eq.~(\ref{501}) gives $(w L)/(2\sqrt{n}) \leq w L \sinh{(\alpha)}) /(2\sqrt{n}) < 1 $.
At the same time, the possibility of accelerated tunneling transitions with respect to the traversal
{\em classical} course, i. e. $t_{T, \varphi}^{(\alpha)}< \tau_{\k}$, occurs only when $\alpha/2 \geq (\alpha/2) \tanh{(\alpha/2)} > 1$.
Since $\alpha = w L\sqrt{1-n}$, the intersection of the ``weak version'' of both of the above constraints, $(w L)/(2\sqrt{n}) < 1$ and $\alpha/2 > 1$,
leads to $n > 2$, which definitely does not correspond to a tunneling configuration.
This analysis leads to a very important conclusion: in the region where the one-way direction 
transmission coefficient prevails over the reflection coefficient, the wave packet transmitted components
should tunnel with a retarded velocity with respect to the classical velocity since, in this case, $t^{(\alpha)}_{T, \varphi}> \tau_{\k}$.
The supposed accelerated transit of the tunneling wave packet, and therefore, superluminal velocities and the Hartman effect,
would never occur for $|T|^{\2} > |R|^{\2}$.

To summarize, we have reported about a way of comprehending the conservation of probabilities \cite{Ber04, Ber06} for a very particular tunneling configuration where the asymmetry and the distortion aspects presented in the standard case were all eliminated.
As a result the phase time could be accurately calculated in order to give the {\em exact} position of the scattered wave packets (or particles), for which, in the standard case, the position of the peak is shifted.
Essentially, we have claimed for relevance of the use of the multiple peak decomposition technique 
in obtaining the exact relation between two distinct scattering/tunneling time definitions: the phase time and the dwell time.
In spite of quoting the superluminal interpretation, our discussion concerned with the definition of the strict mathematical conditions stringing the applicability of the stationary phase principle in deriving transit times.
Even with the introduced modifications, our results corroborate with the analysis \cite{WinWin,Win03} that gives an answer to the paradox of the Hartman interpretation \cite{Har62}.
In certain sense, we accept the fact that one should keep in mind that the Hartman Effect, even in its more sophisticate consequences)
appears to have been experimentally verified \cite{Exp}, in particular, for opaque barriers and nonresonant tunneling \cite{Zai05} and also reproduced by numerical simulations and constrained theoretical analysis \cite{Bar02,Pet03}.
In general lines, there have also been some trying of yielding complex time delays for tunneling analysis, ultimately due to a complex propagation constant where the supposition of superluminal features is considered artificial since the transmitted peak is not causally related to the corresponding incident peak.
This semi-classical method makes use of complex trajectories which, in its turn, enables the definition of real traversal times in the complexified phase space \cite{Xav97,Sok90,Sok94}.
Yet concerning the subsequent theoretical perspectives in the above frameworks, the symmetrical colliding configuration also offers the possibility of exploring some problems involving soliton structures.
In particular, all the above arguments suggest that the idea of complexifying time should be investigated for some other scattering configurations, which reinforces the more general assertion that the investigation of wave propagation across a tunnel barrier has always been an intriguing subject which is wide open both from a theoretical and an experimental point of view.

{\bf Acknowledgments}
We would like to thank FAPESP (PD 04/13770-0) for the financial support.

\end{document}